\documentclass[11pt,a4paper]{article}

\usepackage[utf8]{inputenc}
\usepackage[T1]{fontenc}
\usepackage[margin=2.5cm]{geometry}
\usepackage{amsmath,amssymb}
\usepackage{graphicx}
\usepackage{float}
\usepackage{booktabs}
\usepackage{tabularx}
\usepackage{multirow}
\usepackage{array}
\usepackage{enumitem}
\usepackage[numbers]{natbib}
\usepackage[colorlinks=true,linkcolor=blue,citecolor=blue,urlcolor=blue]{hyperref}

\newcolumntype{L}{>{\raggedright\arraybackslash}X}
\newcolumntype{C}{>{\centering\arraybackslash}X}

\newcommand{\FSTC}{\mathrm{FSTC}}
\newcommand{\FT}{\mathrm{FT}}
\newcommand{\bin}{\mathrm{bin}}

\title{A Quantum Computing Framework for VLBI Data Correlation}

\author{Lei Liu$^{1,*}$ \\[6pt]
\small $^{1}$ Shanghai Astronomical Observatory, Chinese Academy of Sciences, Shanghai 200030, China\\
\small $^{*}$ liulei@shao.ac.cn}

\date{}

\begin{document}

\maketitle

\begin{abstract}
This work presents a quantum computing framework for Very Long Baseline Interferometry (VLBI) data correlation. It is shown that a classical baseband time series data of length N can be embedded into a quantum superposition state using amplitude encoding with only $\log_2 N$ qubits. The basic VLBI correlation and fringe fitting operations, including fringe rotation, Fourier transform, delay compensation, and cross correlation, can be implemented via quantum algorithms with significantly reduced computational complexity. A single-baseline quantum pipeline for geodetic VLBI data processing is constructed, and its feasibility and accuracy are validated through direct comparison with a classical pipeline. Amplitude encoding of large data volumes remains the primary bottleneck in quantum computing. However, the quantized nature of VLBI raw data helps reduce the state-preparation complexity. This investigation demonstrates that quantum computation offers a promising paradigm for VLBI data correlation and is likely to play a role in future VLBI systems.

\vspace{6pt}
\end{abstract}

\section{Introduction}
The concept of a quantum computer was first proposed by Richard Feynman in 1981 at the First Conference on the Physics of Computation \citep{Feynman1982}. Inspired by Feynman's insight, \citet{Deutsch1985} pointed out that, by taking quantum parallelism into account, a quantum computer could solve certain computational problems faster than classical ones.
\citet{Shor1994} reformulated the integer factorization problem as period finding problem and demonstrated that the quantum Fourier transform (QFT) enables period extraction in polynomial time, therefore achieving exponential speedup compared with classical algorithms. 

Since then, quantum computing (QC) developed rapidly \citep{QCQI}. Quantum computers demonstrate quantum advantage in tasks such as Random Circuit Sampling \citep{google} and Gaussian Boson Sampling \citep{jiuzhang}. Scientists across multiple disciplines have begun exploring the potential applications of quantum algorithms, including quantum chemistry, combinatorial optimization, quantum machine learning (QML), etc. \citep{AY2024}. This work focuses on the possible application of quantum algorithms in radio astronomy data processing, which involves superdense coding, data searching and QFT (Quantum Fourier Transform). However, present-day quantum computers are still subject to NISQ (Noisy Intermediate-Scale Quantum)-era limitations. Notably, amplitude encoding is still a non-trivial problem for interferometric data, preventing the practical implementation of the corresponding quantum processing pipelines as we will introduce in this work.

Very Long Baseline Interferometry (VLBI) is the astronomical technique that achieves the highest angular resolution among all direct imaging techniques \citep{VLBI}, and has been widely used in astrophysics \citep{EHT}, astrometry \citep{VGOS}, and orbit determination of spacecraft \citep{Duev2012}. Owing to its operation in the radio-frequency domain, VLBI relies heavily on digital signal processing, including data recording \citep{Whitney2007}, channelization \citep{Zhu2016}, and correlation \citep{CVNCORR, difx}. 

It is worth mentioning that interferometric correlation requires access to the original baseband data. Even with 1-bit or 2-bit quantization, the resulting data volume remains large. Moreover, the baseband data are actually white noise, rendering classical data compression techniques ineffective \citep{VLBI}. Also note that in VLBI data processing, the computational load of cross-correlation part scales quadratically with the number of stations. To meet the challenges posed by the rapidly growing data volumes and computational demands of next-generation facilities, dedicated data processing infrastructures are required. An example is the SKA (Square Kilometre Array) \citep{SKA} Regional Centre (SRC) network, which will handle the distribution, processing, and storage of 700 PB data per year.

This work shows that quantum architecture is well suited for radio interferometric data storage and processing. First of all, a system of $n$ qubits spans a Hilbert space of dimension $2^n$, enabling exponentially large classical data to be represented. As a result, a classical signal of length $N$ can be embedded into a quantum superposition state using amplitude encoding, requiring only $\log_2 N$ qubits and achieving an impressive form of ``data compression''. Moreover, linear phase manipulations, such as time-domain fringe rotation or fractional sample time correction (FSTC) in frequency-domain, can be implemented via quantum linear operators using binary decomposition, where each qubit receives a single phase rotation. This reduces the computational complexity from $O(N)$ to $O(\log_2 N)$. The time to frequency domain transform can also be executed efficiently through QFT, which operates with complexity $O((\log_2 N)^2)$ \citep{QCQI}. Finally, while quantum circuits are not suitable for computing cross-power spectrum directly, the correlation between two spectrum represented as multi-qubit quantum states naturally corresponds to their inner product. Within the quantum framework, this inner product can be evaluated by constructing a unitary composite operator and extracting the result using Hadamard test \citep{QCQI}. The effect of residual delay on the cross-power spectrum can be embedded into this process, which makes post-processing fringe fitting possible. This workflow is particularly well suited for geodetic VLBI, where the extraction of delay and delay-rate observables is essential \citep{Kondo2016}.

Above analysis indicates that the construction of a quantum-based correlator is theoretically sound and technically viable. 
In this work, by implementing all the necessary quantum operations and building a single-baseline geodetic VLBI data correlation pipeline using \texttt{Qiskit} \citep{Qiskit}, the feasibility of utilizing quantum computing for VLBI data processing at theoretical and algorithmic level is demonstrated. Moreover, the advantages of computational complexity offered by the quantum algorithms compared to their classical counterparts are exhibited. 
It should be noted that the current pipeline remains primarily applicable to geodetic VLBI data processing. However, the large data volumes expected from next-generation VLBI arrays could potentially benefit from a quantum computing approach. Further development and optimization of the architecture will be necessary to satisfy requirements of radio astronomical observations.

This paper is organized as follows: Section~\ref{sec:framework} introduces the whole quantum computing framework. Section~\ref{sec:pipeline} presents the quantum pipeline and performs comparison with classical pipeline. Section~\ref{sec:discussion} and Section~\ref{sec:conclusion} are discussion and conclusion.

\section{The Quantum Computing Framework}\label{sec:framework}

\begin{figure}[H]
    \centering
    \includegraphics[width=\linewidth]{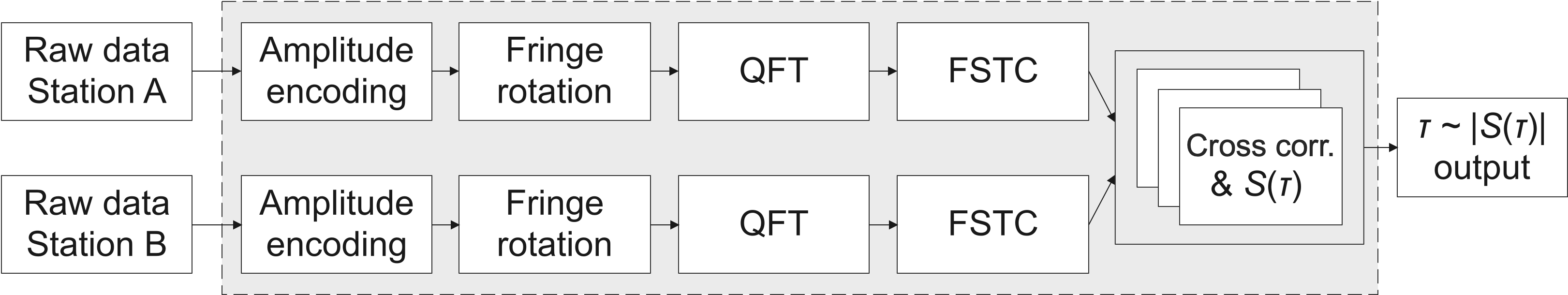}
    \caption{Dataflow of quantum pipeline. Quantum circuits are highlighted with gray background.\label{fig:flow_quantum}}
\end{figure}

\begin{figure}[H]
    \centering
    \includegraphics[width=\linewidth]{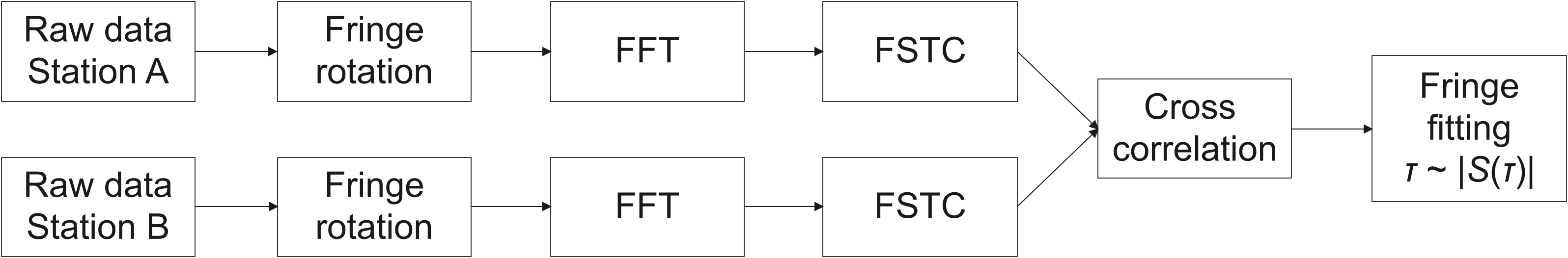}
    \caption{Dataflow of classical pipeline.\label{fig:flow_classic}}
\end{figure}

The quantum framework proposed in this work involves all the necessary quantum operations required for building the VLBI data correlation pipeline. The dataflow of the quantum pipeline is presented in Figure~\ref{fig:flow_quantum}. For comparison, the classic pipeline is also presented (Figure~\ref{fig:flow_classic}). 

For both pipelines, the data processing procedures are similar:
\begin{itemize}
\item Raw data loading: raw data from the two stations (A and B in this work) forming a single baseline are loaded into the data processing pipeline.
\item Fringe rotation: a time-domain frequency shift that fully aligns the signal frequencies from the two stations.
\item Fourier transform: conversion of signal from time domain to frequency domain.
\item Fractional sample time correction (FSTC): an operation in frequency domain that fully aligns the signal delays from the two stations.
\item Correlation: conjugate multiplication of corresponding frequency bins from the two stations.
\item Fringe fitting: searching for non-zero residual delay and delay rate within the cross spectrum to achieve a fully flattened fringe phase.
\end{itemize}
This section presents implementation of above operations in the quantum computing framework. Throughout this work, the raw baseband data are real-valued signals. Given that fringe rotation, FSTC, and fringe fitting are all based on linear phase modulation of a given sequence, they are treated collectively in Section~\ref{sec:phase_mod}.

To make the framework easy to understand, the code associated with this work is available in the GitHub repository\footnote{\url{https://github.com/liulei/qcorr}}, with annotations referencing the corresponding equations in this paper. Readers are encouraged to cross-reference the code with the text.

\subsection{Data Encoding}
\begin{figure}[H]
    \centering
    \includegraphics[width=0.8\linewidth]{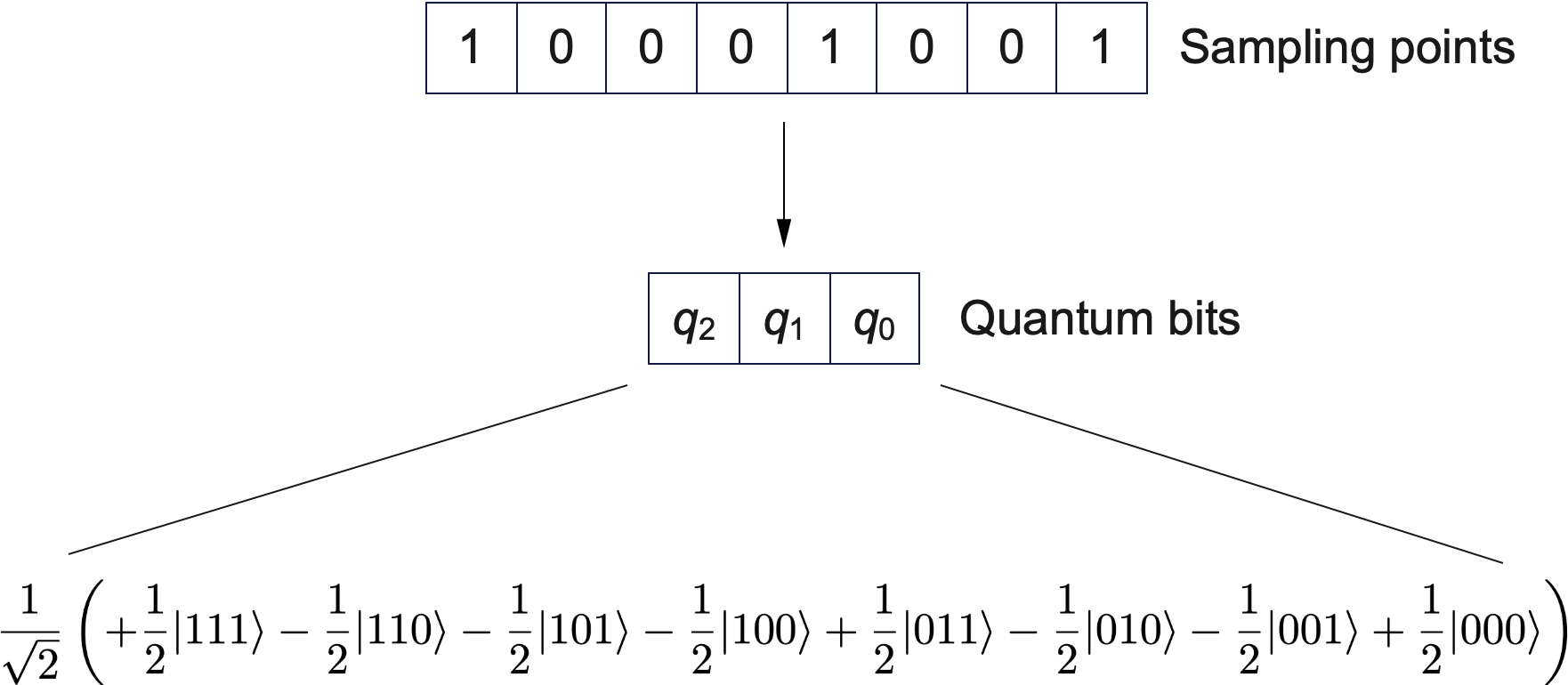}
    \caption{Demonstration of amplitude encoding. The 8 sampling points with 1-bit quantization are embedded into 3 qubits quantum superposition state via amplitude encoding. The $\frac{1}{\sqrt{2}}$ outside the parentheses is the normalization factor. \label{fig:encoding}}
\end{figure}

In classical data correlation pipeline, the voltage-sampled time series are first loaded from disk into memory and then converted from quantized bit to floating-point representations, e.g., mapping 1-bit quantization levels $\{0,~1\}$ to $\{-\frac{1}{2},~+\frac{1}{2}\}$. In quantum framework, a system of $n$ qubits spans a Hilbert space of dimension $2^n$, allowing exponentially large classical data to be represented using only a small number of qubits. In other words, a classical signal of length $N$ after normalization can be embedded into quantum superposition state and requires only $n = \log_2 N$ qubits:
\begin{equation}
    |\psi_x\rangle = \sum_{k=0}^{N-1} x_k |k\rangle, 
\end{equation}
where $x_k$ is the normalized value of the $k$-th sampling point:
\begin{equation}
\sum_{k=0}^{N-1} |x_k|^2=1.
\end{equation}
If the number $k$ is expanded in its binary representation:
\begin{equation}\label{eq:k_binary}
    k = \sum_{i=0}^{n-1}2^i k_i, 
\end{equation}
then $|k\rangle=|k_{n-1} \dots k_1 k_0\rangle$ is the $k$-th state of the quantum system composed of $n$ qubits (Figure~\ref{fig:encoding}). 

Amplitude encoding is used to load raw data into quantum state. It is realized that this is a huge challenge to the current quantum circuits. The possible solution will be discussed in Section~\ref{sec:diss_encoding}.

\subsection{Phase Modulation}\label{sec:phase_mod}
\begin{figure}[H]
    \centering
    \includegraphics[width=0.5\linewidth]{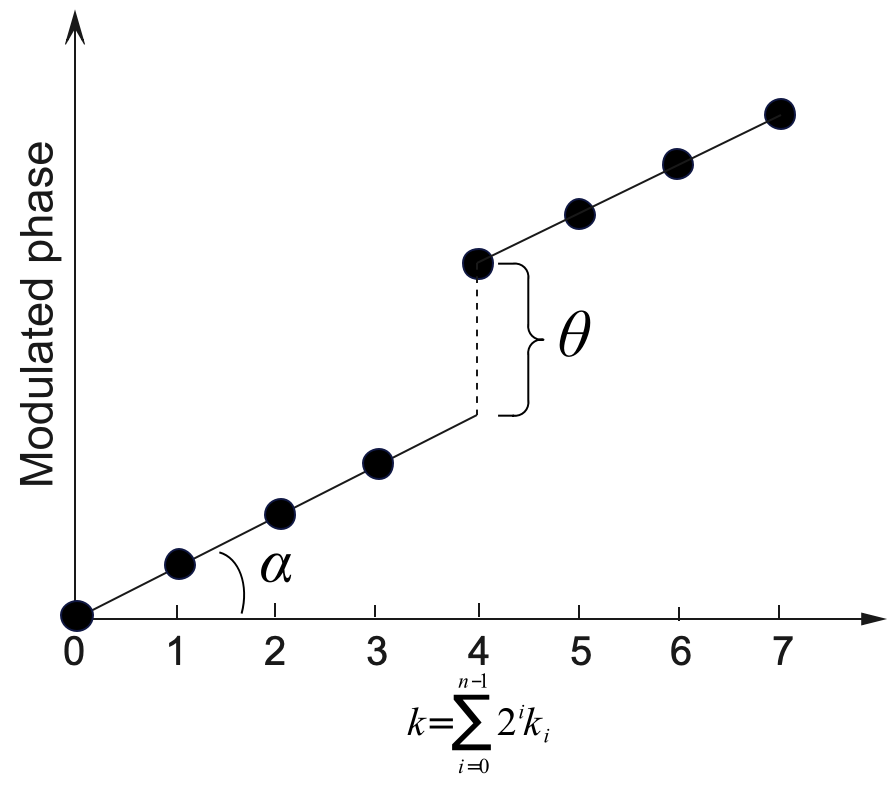}
    \caption{Demonstration of phase modulation. $\alpha$ corresponds to the slope of the linear phase modulation in Equqation~\ref{eq:linear_ph_mod}. $\theta$ denotes the global phase shift applied to the ($n-1$)-th qubit in Equation~\ref{eq:n1_ph_mod}.\label{fig:phase_mod}}
\end{figure}

For VLBI data correlation and post processing, phase modulations for the given sequence are required in several scenarios:
\begin{itemize}
    \item Fringe rotation in the time domain: 
\begin{equation}\label{eq:fringe_rot}
x_k \rightarrow x_k e^{-j2\pi\Delta f T_s k},
\end{equation}
where $\Delta f=f_0\dot{\tau}$ is the fringe rate determined by the sky frequency $f_0$ and delay rate $\dot{\tau}$ given by the delay model. $T_s=1/f_s$ is the sampling time, which is proportional to inverse of sampling rate $f_s$.
    \item Fractional Sample Time Correction (FSTC):
\begin{equation}\label{eq:fstc}
s_k\rightarrow s_k e^{-j2\pi f_\mathrm{bin}\tau_\mathrm{frac} k},
\end{equation}
where $f_\mathrm{bin}$ is the frequency bin width in cross correlation. $\tau_\mathrm{frac}$ is the FSTC value given by the delay model. 
    \item Fringe fitting in post processing: 
\begin{equation}\label{eq:fringe_fit}
    p_k\rightarrow p_k e^{-j2\pi f_\mathrm{bin} \tau_\mathrm{residual}k},
\end{equation}
where $\tau_\mathrm{residual}$ is the residual delay. 
\end{itemize}
Note that above operations are mathematically identical, and the modulations are linear, which can be expressed as: 
\begin{equation}
x_k\rightarrow x_k e^{j\alpha k}.\label{eq:linear_ph_mod}
\end{equation}
In classical signal processing, phase modulation for a sequence of length $N$ is pointwise, which yields a computational complexity of $O(N)$. However, in the quantum computation framework, the linear phase modulation can be represented by the unitary operator (Figure~\ref{fig:phase_mod}):
\begin{equation}\label{eq:Ualpha}
P(\alpha):~|k\rangle \rightarrow e^{j\alpha k} |k\rangle.
\end{equation}
The corresponding phase factor for $n$ qubits can be written as:
\begin{equation}
    e^{j\alpha k} = \prod_{i=0}^{n-1} e^{j\alpha 2^i k_i},
\end{equation}
which means the phase modulation can be implemented as a phase rotation gate $P(\alpha 2^i)$ for the $i$-th qubit. Since quantum gates act globally on all components of a superposed quantum state, this operation modifies amplitudes of all $N$ basis simultaneously. The computational complexity is reduced from $O(N)$ of classic operations to $O(n)=O(\log_2 N)$ of quantum operations.

It should be noted that the linear approximation required for the linear phase modulation in this section imposes an upper limit on $N$. The maximum allowable integration time by assuming the corresponding term in the delay model polynomial is neglected, is presented in Table~\ref{tab:Tmax}. Based on this table and the sampling time, the upper limit of $N$ can be estimated.

\subsection{Fourier Transform}
In VLBI data correlation, fringe-rotated baseband samples are Fourier transformed into the frequency domain for subsequent FSTC and cross-correlation. Classically, this step relies on the Fast Fourier Transform (FFT), which has a computational complexity of $O(N \log_2 N)$. In the quantum framework, an analogous operation is provided by the Quantum Fourier Transform (QFT). For an $n$ qubit quantum state, the QFT acts on all $2^n$ amplitude components simultaneously through a sequence of Hadamard gates and controlled-phase rotations, achieving an overall complexity of $O(n^2)$. 

\subsection{Cross Correlation}\label{sec:Ucorr}
This work notes that quantum circuits are actually not suitable for direct output of full cross-power spectrum between two signals, since this involves the extraction of visibility in each individual frequency bin. The corresponding computational complexity is $O(N)$. Instead, when radio interferometric signals are encoded as quantum states of $n$ qubits for each station, the cross correlation between the two stations is naturally the inner product of the corresponding quantum states.

Assuming $|\psi_A\rangle$ and $|\psi_B\rangle$ are Fourier transformed spectrum of station A and B, which can be expressed as 
\begin{equation}
|\psi_A\rangle = \sum_{k=0}^{N-1}s_{a, k}|k\rangle,\qquad |\psi_B\rangle = \sum_{k=0}^{N-1}s_{b, k}|k\rangle,
\end{equation}
where $s_{a,k}$ and $s_{b,k}$ are the normalized complex spectrum at the $k$-th frequency point. The classical cross correlation, which is expressed as the sum of point-wise complex multiplications:
\begin{equation}
S = \sum_{k=0}^{N-1} s_{a,k} s_{b,k}^\ast,
\end{equation}
is precisely the quantum mechanical inner product:
\begin{equation}
S =\mathcal{M}( \langle \psi_B | \psi_A \rangle)
\end{equation}
Therefore, the cross correlation computation, or the coherent summation of cross spectrum across all frequency bins, is implicitly embedded into the multi-qubit Hilbert space and does not require evaluating each term $s_{a,k} s_{b,k}^{\ast}$ individually. Note that the inner product is in quantum state and must be ``read out'' through a specific measurement $\mathcal{M}$. This work notes that the measurement is experimentally not direct: the complex inner product cannot be observed directly. In the quantum computing framework, its value can be extracted from the probability distribution of repeated ancilla measurements via Hadamard test. Let
\begin{equation}\label{eq:U}
U = U_B^\dag U_A
\end{equation}
be the composite operator such that:
\begin{equation}
U_A |0\dots 0\rangle = |\psi_A\rangle,\qquad
U_B |0\dots 0\rangle = |\psi_B\rangle,
\end{equation}
then 
\begin{equation}
\langle \psi_B | \psi_A \rangle=\langle 0\dots 0 | U | 0\dots 0 \rangle.
\end{equation}
The real and imaginary parts of above quantity can be obtained from single-qubit measurement statistics of the ancilla:
\begin{equation}
\mathrm{Re}(\langle \psi_B | \psi_A \rangle)
= 2~\mathrm{Re}(P_0) - 1,\qquad 
\mathrm{Im}(\langle \psi_B | \psi_A \rangle)
= 2~\mathrm{Im}(P_0) - 1,
\end{equation}
where $\mathrm{Re}(P_0)$ and $\mathrm{Im}(P_0)$ denote the probabilities of measuring the ancilla in state $|0\rangle$ for the real and imaginary part of Hadamard tests, respectively.
The detailed derivation of the relationship between $P_0$ and the real/imaginary parts of the inner product in the Hadamard test, as well as the specific implementation of the measurement, are provided in Appendix~\ref{sec:Hadamard}.

As a result, the classical cross correlation operation, which requires a computational complexity of $O(N)$, is replaced by a single quantum inner-product extraction that acts on the entire $2^{n}$-dim superposition state with a computational complexity of $O(1)$. 

Note that when the original baseband data is real signal, for $N$ point real-sampled data, only the first $N/2$ frequency bins are utilized after transforming to the frequency domain. Within the current quantum framework, this corresponds to extracting the coherently summed value of the cross-power spectrum across the corresponding frequency bins. Since the cross-power spectrum is represented as a quantum superposition state within this framework, the first half of the spectrum corresponds to the subspace where the most significant qubit is in the $|0\rangle$ state. As a result, a quantum circuit can be designed to apply a phase rotation operation to the $(n-1)$-th qubit (Figure~\ref{fig:phase_mod}):
\begin{equation}
    D(\theta): |k_{n-1}\dots k_1 k_0\rangle \rightarrow e^{j\theta k_{n-1}}|k_{n-1}\dots k_1 k_0\rangle. \label{eq:n1_ph_mod}
\end{equation}
This operation can be incorporated into the composite operator defined in Equation~(\ref{eq:U}):
\begin{equation}\label{eq:Utheta}
U(\theta) = U_B^\dag D( \theta) U_A,
\end{equation}
Based on the binary representation of $k$ in Equation~(\ref{eq:k_binary}), the summed cross-power spectrum $S$ as a function of $\theta$ can be expressed as:
\begin{equation}
    S(\theta) = \sum_{k=0}^{N/2-1}s_{a,k}s_{b,k}^* + \sum_{k=N/2}^{N-1}e^{j\theta} s_{a,k}s_{b,k}^* = S_0 + e^{j\theta}S_1.
\end{equation}
The required component $S_0$ can be obtained by performing measurements at $\theta=0$ and $\theta=\pi$:
\begin{equation}
    S_0 = (S(0) + S(\pi)) / 2.
\end{equation}

\subsection{Fringe Fitting}\label{sec:fit}
Fringe fitting is an important step in VLBI data post processing. As a first demonstration of applying quantum computation to VLBI data correlation, a simplified scenario in which a single IF (Intermediate Frequency) is used to estimate a single residual delay parameter $\tau$. The corresponding fitting function $S(\tau)$ is given by:
\begin{equation}\label{eq:Stau}
    S(\tau)=\sum_{k=0}^{N-1}s_{a,k} s_{b,k}^\ast e^{-j2\pi f_\mathrm{bin}\tau k}.
\end{equation}
The purpose of fringe fitting is to find out appropriate $\tau$ that maximizes $|S(\tau)|$. In actual application, fringe fitting is a generalization of this simplified case, while the underlying principle remains the same.

For the quantum circuits in this work, the construction of the above fitting function corresponds to incorporating a linear phase modulation associated with the delay search into the composite operator defined in Equation~(\ref{eq:U}):
\begin{equation}\label{eq:Utau}
U(\tau) = U_B^\dagger P(2\pi f_\mathrm{bin}\tau)U_A.
\end{equation}
The summed value of the phase-modulated cross-power spectrum can be written as
\begin{equation}
\langle \psi_B | \psi_A \rangle (\tau)=\langle 0\dots 0 | U(\tau)| 0\dots 0 \rangle, 
\end{equation}
in its quantum state and then be extracted to obtain the fitting function:
\begin{equation}
S(\tau)=\mathcal{M}(\langle \psi_B | \psi_A \rangle (\tau)).
\end{equation}

Fringe fitting can be regarded as a global optimization problem, in which the residual delay serves as the fitting parameter and the fitting function is the objective function. When the correct residual delay is identified, the fringe phase becomes flattened, and the objective function reaches its maximum. Within the present quantum algorithmic framework, the fitting function $S(\tau)$ is evaluated for a series of trial delay $\tau$ using the quantum procedure described above.

In our quantum algorithmic framework, the cross-correlation and fringe-fitting procedures are actually coupled together. In practice, one first determines a set of equally spaced trial delays based on the search range and step size. Then for each trial value $\tau$, the composite operator $U(\tau)$ is constructed following the method described above, and the corresponding $S(\tau)$ is obtained via Hadamard test. The residual delay is then identified as the value of $\tau$ at which $|S(\tau)|$ reaches its maximum. 

It should be noted that, since $\tau$ must be discretized in this procedure, there is a discrepancy between the discrete and the true maximum. Therefore, a further polynomial fit of the $\tau\sim|S(\tau)|$ relation is usually required to further improve the estimation of the true maximum. This step is currently carried out classically and will be introduced in Section~\ref{sec:polyfit}.

This section presents only a fundamental framework for implementing a quantum computing based correlator. Operations required for processing actual observation data, such as integration and delay rate searching, can in principle be incorporated within this framework, and will be discussed further in Sections~\ref{sec:integration} and \ref{sec:fringe_fitting}.

\section{Demonstration of Quantum Processing Pipeline}\label{sec:pipeline}

\begin{figure}[H]
    \centering
    \includegraphics[width=0.85\linewidth]{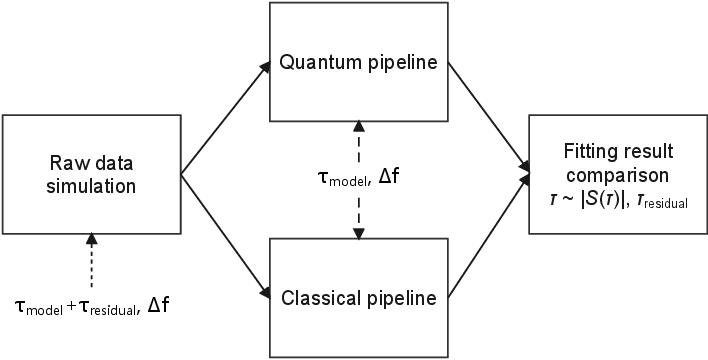}
    \caption{Architecture for the comparison between quantum and classical pipelines.\label{fig:flow_compare}}
\end{figure}

As the demonstration of above quantum based framework, in this section, the complete quantum VLBI data correlation pipeline is constructed. Correlation and fringe fitting are performed with this pipeline, and the delay estimates derived from the quantum pipeline are compared with those obtained from the classical pipeline.

\subsection{Architecture of the System}
To validate the quantum algorithm proposed in this work, in addition to the quantum pipeline itself, a raw-data simulation module for testing, a classical pipeline corresponding to the quantum algorithm, and a classical polynomial fitting module to achieve higher accuracy for residual delay fitting are also developed. The overall architecture is shown in Figure~\ref{fig:flow_compare}.

\subsubsection{Raw Data Simulation}

The parameters for raw data setup are listed in Table~\ref{tab:param_simu}. The input fringe rate and delay are used for raw data generation. The model fringe rate and delay are adopted for frequency and delay correction in the correlation process. The residual delay is kept as the reference value for the verification of the two pipelines. Two streams of raw data are generated based on a Gaussian distribution, each containing $N=256$ unquantized sampling points. After that, an input delay and fringe rate are applied. Although no explicit noise is added during the entire process, numerical noise is inevitably introduced during delay insertion, fringe rotation, and real signal reconstruction stages.
Current processing implements cross correlation for only one Fourier transform period, with no further integration applied. By normalizing the baseline cross-power spectrum using the method from HOPS (Haystack Observatory Postprocessing System) \citep{HOPS}, an SNR (Signal-to-Noise Ratio) of 11.9 is achieved, which is calculated based on the normalized amplitude and the corresponding bandwidth and integration time.

\subsubsection{Quantum Pipeline}
The quantum pipeline is implemented with \texttt{Qiskit}. Classical raw data of length $N$ are embedded into quantum system with $n=\log_2N$ qubits via amplitude encoding. The linear phase modulation operators for fringe rotation and FSTC are constructed based on Equation~(\ref{eq:Ualpha}). Besides that, this operator is also used in the construction of composite operator $U(\tau)$. The transform from time domain to frequency domain is performed with QFT. The inner-product of two quantum spectrum are extracted via Hadamard test. To support fringe fitting, for each trial residual delay $\tau$, the composite operator $U(\tau)$ is constructed according to Equation~(\ref{eq:Utau}). 

\subsubsection{Classical Pipeline} 
In classical VLBI data correlation pipeline, the whole process can be divided into two stages. The first stage is cross correlation, including fringe rotation, FFT, FSTC and cross correlation. The cross spectrum is saved for further analysis. The second stage is post processing: for each tentative delay $\tau$, $S(\tau)$ is calculated according to Equation~(\ref{eq:Stau}). The differences between quantum and classical pipeline are clearly demonstrated in Figures~\ref{fig:flow_quantum} and \ref{fig:flow_classic}. In our current implementation, the fringe rotation, FSTC and delay compensation are implemented according to Equations~(\ref{eq:fringe_rot}), (\ref{eq:fstc}) and (\ref{eq:fringe_fit}). The Fourier transform is performed with standard FFT.

\begin{figure}[H]
    \centering
    \includegraphics[width=0.8\linewidth]{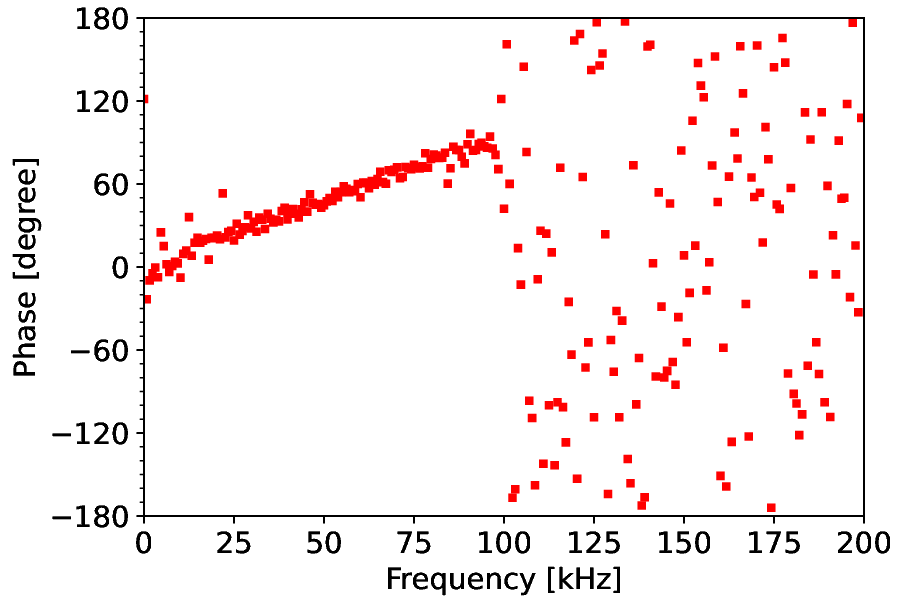}
    \caption{Demonstration of the cross spectrum after correlation. Red square represents phase value at the corresponding frequency bin. For real signal data used for pipeline verification in this work, only the first half frequency bins contain useful information. Since delay compensation is not performed, the fringe is not flat. The fitting of the slope yields the residual delay listed in Table~\ref{tab:param_simu}.\label{fig:spec_tau0}}
\end{figure}

\subsubsection{Polynomial Fit \& Comparison}\label{sec:polyfit}

In VLBI data post processing, polynomial fit for the discretized values of $|S(\tau)|$ is usually performed, so as to achieve a better estimation of the residual delay. The procedure is:
\begin{itemize}
    \item Identify the maximum amplitude and the corresponding delay from a series of equally spaced trial delays for the given search range and step size. 
    \item Perform second order polynomial fit for the consecutive 5 points around the maximum amplitude, with $x$ and $y$ correspond to $\tau$ and $\vert{}S(\tau)\vert{}$, respectively.
    \item Derive $\tau$ ($x$) that corresponds to the maximum value of the polynomial curve. If required, also derive the uncertainty of delay $\sigma_\tau$ based on the covariance matrix. 
\end{itemize}

The $\tau\sim|S(\tau)|$ relations from both pipelines with identical raw data as input are fitted and compared, so as to evaluate the data correlation accuracy of the quantum pipeline. The raw cross-spectrum before fringe fitting is dumped and plotted in Figure~\ref{fig:spec_tau0}. The slope of the fringe phase corresponds to the pre-set residual delay. Since our simulated data are real signal, only the first half frequency bins are adopted. In the quantum pipeline, the summed cross spectrum of this part is extracted via Hadamard test as explained in Section~\ref{sec:Ucorr}.

\subsection{Simulation Result}

\begin{table}[H]
\caption{Raw data setup for quantum pipeline verification.\label{tab:param_simu}}
\begin{tabularx}{\textwidth}{LL}
\toprule
Parameter   &  Value \\
\midrule
Data type   &  Real signal, unquantized \\
Sampling number &   256 \\
Qubit number    &   8   \\
Bandwidth       &   0.1~MHz \\
Input/model fringe rate ($\Delta f$)    &   500~Hz \\
Input delay ($\tau_\mathrm{model}+\tau_\mathrm{residual}$)  & 6.7~$\mu$s \\
Model delay ($\tau_\mathrm{model}$)  & 4.0~$\mu$s \\    
Residual delay ($\tau_\mathrm{residual}$)  & 2.7~$\mu$s \\
\bottomrule
\end{tabularx}
\end{table}

\begin{figure}[H]
    \centering
    \includegraphics[width=0.8\linewidth]{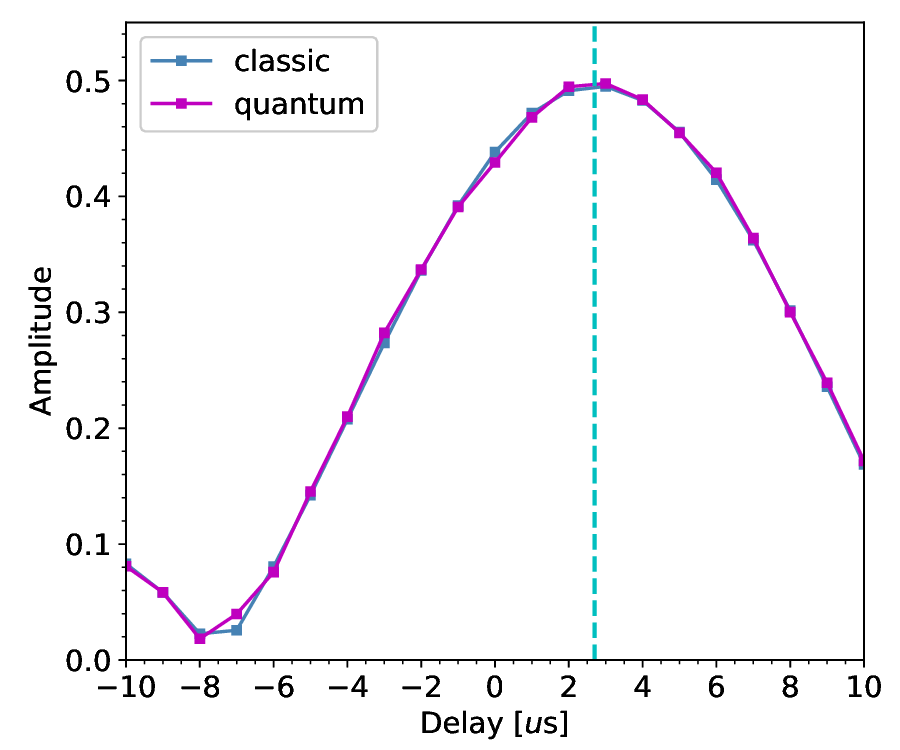}
    \caption{The amplitude of the summed cross-spectrum after delay compensation $|S(\tau)|$ as a function of trial delay $\tau$. Cyan dashed line represents the input residual delay $\tau_\mathrm{residual}=2.7~\mu\mathrm{s}$. 
Blue line corresponds to the fitting result of classic pipelines, with the derived residual delay and uncertainty of $2.734\pm0.003~\mu$s. The quantum pipeline generates multiple sets of results under various combinations of shot number and delay search resolution. A representative case with shots = 20000 and resolution = 1.0~$\mu$s is illustrated by the magenta line. More fitting results for different configurations are presented in Table~\ref{tab:shots_res}.\label{fig:polyfit}}
\end{figure}

Figure~\ref{fig:polyfit} demonstrates the $\tau\sim|S(\tau)|$ relationships as the output of both classic and quantum pipelines. In general, the amplitudes of the two pipelines agree well with each other. Each pipelines derive a residual delay that is close to the real value. 
For fringe fitting within the current quantum computing framework, a critical issue is the impact of shot noise and delay search resolution on measurement accuracy. To address this, the estimated delay values and their corresponding uncertainties ($\sigma$) for the combinations of various shot numbers (5000, 20000 and 40000) and delay search resolutions ($1~\mu\text{s}$, $0.5~\mu\text{s}$ and $0.25~\mu\text{s}$) are evaluated. To make the results more reliable, this work runs the quantum pipeline 10 times for each specific combination of shots and resolution, and then calculates the mean measured value and $\sigma$. The results are shown in Table~\ref{tab:shots_res}. As expected, a larger number of shots corresponds to a smaller $\sigma$, indicating that increasing the shot count effectively mitigates the impact of shot noise. Moreover, as the resolution improves from $1~\mu\text{s}$ to $0.25~\mu\text{s}$, the corresponding $\sigma$ decreases, indicating that the resolution utilized during curve fitting indeed significantly affects the estimation precision. It should be noted that the resolution cannot be arbitrarily high within present delay search algorithm, as this would consume too much quantum computing resources. A more ``QC-style'' approach is to modify current fringe fitting method to make it better suited for quantum computation, which is discussed in Section~\ref{sec:fringe_fitting}. On the other hand, although the $\sigma$ obtained from the classical pipeline is much higher than that of the quantum pipeline, its measured value still exhibits a discrepancy from the actual value of $2.7~\mu\text{s}$ within the $3\sigma$ intervals. This implies that this formal error is not as representative as it appears. This work acknowledges that applying quantum computing to the data processing of radio interferometry signals remains an emerging field, and the exact influence of quantum measurement noise on the final results still requires further comprehensive analysis.

\begin{table}[H]
\caption{Estimated delay values and uncertainties ($\sigma$) for combinations of shot numbers and delay search resolutions. All values are presented in units of $\mu$s.\label{tab:shots_res}}
\begin{tabularx}{\textwidth}{CCCC}
\toprule
\multirow{2}{*}{{Resolution ($\mu$s)}} & \multicolumn{3}{c}{{Shot numbers}} \\
\cmidrule{2-4}
& {5000} & {20000} & {40000} \\
\midrule
1.00 & 2.730$\pm$0.243 & 2.811$\pm$0.153 & 2.767$\pm$0.059 \\
0.50 & 2.748$\pm$0.167 & 2.706$\pm$0.117 & 2.740$\pm$0.054 \\
0.25 & 2.805$\pm$0.159 & 2.657$\pm$0.074 & 2.717$\pm$0.048 \\    
\bottomrule
\end{tabularx}
\end{table}

\section{Discussion}\label{sec:discussion}
It must be noted that current algorithm design and hardware implementation are not yet ready to support a deployable quantum correlation pipeline. In this section，the associated challenges and their possible solutions are discussed. These will also be the follow-up work of this study.

\subsection{Amplitude Encoding}\label{sec:diss_encoding}
In this work, amplitude encoding is adopted to load classical raw data into quantum superposition states. However, given the current capabilities of quantum circuits, preparing an arbitrary quantum state remains a major challenge even in the algorithm level. A computational complexity of $O(N)$ for encoding $N$ samples makes it the most expensive component of the entire quantum framework. Actually data loading is widely recognized as the main bottleneck that of many quantum algorithms \citep{AY2024}. 

Notice that VLBI baseband data are always recorded in quantized form, e.g., in the 1-bit case, all samples have the same magnitude and differ only in sign, and large contiguous regions may contain identical values. By exploiting these patterns and employing a binary-tree structure, in the best case it is possible to reduce the complexity of loading $N$ samples to $\log_2 N$. It is planned to implement this amplitude-encoding scheme in future work and evaluate its performance using real VLBI raw baseband data.

It should be noted that data encoding is not merely an algorithmic challenge. From a hardware perspective, it requires high-fidelity coherent control for the target superposition state, which remains a major bottleneck on current NISQ devices. Therefore, the practical application of quantum computing in a wide range of fields relies on advances in quantum hardware.

\subsection{Cross Spectrum Integration}\label{sec:integration}

\begin{table}[H]
\caption{Maximum integration time assuming the corresponding term in delay model polynomial is ignored. The sky frequency and bandwidth are set to 8.4~GHz and 0.1~MHz, respectively. The phase error threshold is set to 0.1~rad. The integration time is derived using the method described in Appendix~\ref{sec:Tmax}.\label{tab:Tmax}}
\begin{tabularx}{\textwidth}{LCCC}
\toprule
&  {$a_1$ ($\sim10^{-6}~\mathrm{s/s}$)} & {$a_2$ ($\sim10^{-11}~\mathrm{s/s}^2$)} & {$a_3$ ($\sim10^{-15}~\mathrm{s/s}^3$)} \\
\midrule
Fringe rotation &  8~$\mu$s &   1.23~s  &   31.18~s\\
FSTC            &  2~ms     &   19.95~s &   199.65~s\\   
\bottomrule
\end{tabularx}
\end{table}

The quantum algorithm presented in this work demonstrates cross-correlation of only a single Fourier transform (FT) period. In practice, VLBI data correlation requires coherent integration of cross-power spectra over many FT periods, so as to improve the signal-to-noise ratio. 
The most straightforward approach for the implementation of integration is to calculate the modulation coefficients for fringe rotation and FSTC for each FT period based on the delay model, perform the correlation, and then read out the correlation results for that specific FT period via Hadamard test. The integration along the time axis and the extraction of delay and delay rate are both executed on the classical pipeline. By choosing a larger $N$, the advantage of the quantum pipeline in terms of computational complexity can be exploited, and the number of Hadamard test readouts for correlation results can be reduced. However, $N$ cannot be infinitely large; otherwise, it would break the linear approximation of phase modulation for a given sequence. E.g., Table~\ref{tab:Tmax} gives the estimation of maximum allowable integration time if the corresponding polynomial term is not taken into account. The corresponding sky frequency $f_0$ and bandwidth $B$ are selected as typical values of 8.4~GHz and 32~MHz, respectively. The phase error threshold $\epsilon$ is set to 0.1~rad. At present, under the configuration of only a single FT period where only linear terms are considered for fringe rotation and FSTC, the maximum allowable integration times are 1.23~s (for $a_1$ only) and 2~ms (for $a_0$ only). After incorporating $a_2$ and $a_1$ into the considerations, the maximum integration times reach 32.18~s and 19.95~s, which are basically the same order as a typical scan length. Therefore, it is essential to consider loading data of multiple FT periods simultaneously into a quantum superposition state, and implement higher-order compensation for fringe rotation and FSTC within the quantum pipeline through an improved algorithm. This approach is more ``quantum computing style'' and efficient, as the readout frequency of the quantum inner product will be substantially reduced. The specific steps are as follows:

\begin{enumerate}[label=\arabic*.]
\item Data loading. Data of multiple FT period are load into quantum superposition state via amplitude encoding. Let $|m,k\rangle$ denote the tensor-product basis corresponding to the $m$-th FT period and the $k$-th sampling point inside this FT period. The quantum state of one station can then be written as:
\begin{equation}
    |\psi\rangle = \sum_{m,k} x_{m,k}| m,k\rangle,
\end{equation}
which composes of $n_m + n_k$ qubits. Here $n_m=\log_2 N_m$ and $n_k=\log_2 N_k$, with $N_m$ and $N_k$ the number of FT periods and sampling points within a FT period, respectively.

\item Fringe Rotation (FR). To support longer integration times, the time-dependent variation of the delay rate must be considered. The delay rate could be updated once per FT period based on the acceleration term $\ddot{\tau}$:
\begin{equation}
    \dot{\tau}(m) = \dot{\tau}_0 + \ddot{\tau} m T_\FT.
\end{equation}
The corresponding phase modulation is:
\begin{align}
    \phi_\mathrm{FR} (m,k) & = 2\pi f_0 \dot{\tau}(m) (kT_s + mT_\FT)\nonumber \\
        & = 2\pi f_0 (\dot{\tau}_0 kT_s+\ddot{\tau}mkT_\FT T_s \nonumber \\
        & ~~~~ +\dot{\tau}_0 mT_\FT+\ddot{\tau}m^2T_\FT^2)
\label{eq:fr_mk}
\end{align}

The phase modulation operator takes the form: 
\begin{equation}\label{eq:P4}
P(\alpha, \beta, \gamma, \eta): |m,k\rangle\rightarrow e^{j(\alpha k +\beta mk + \gamma m + \eta m^2)} |m,k\rangle
\end{equation}
with
\begin{align}
    \alpha & = 2\pi f_0\dot{\tau}_0 T_s \nonumber \\
    \beta  & = 2\pi f_0\ddot{\tau} T_\FT T_s \nonumber \\
    \gamma & = 2\pi f_0\dot{\tau}_0 T_\FT \nonumber \\
    \eta   & = 2\pi f_0\ddot{\tau}T_\FT^2 \nonumber
\end{align}
In Equation~(\ref{eq:P4}), the first and third term correspond to the linear phase modulation within and between the FT period. The second and fourth term represent the coupling between linear phase modulations, which are the second-order non-linear terms. It is noted that these terms can be efficiently decomposed into binary forms by introducing controlled-phase gates. For the $mk$ term, since:
\begin{equation}
m = \sum_p 2^p m_p,~~k = \sum_i 2^i k_i,
\end{equation}
therefore:
\begin{equation}
mk  =   \sum_{p, i} 2^{p+i}m_p k_i.
\end{equation}
The corresponding quantum circuit can be constructed as the controlled-phase gate of $CP(\beta 2^{p+i})$ between each pair of qubits ($m_p, k_i$). This implementation naturally integrates the second-order delay model into the proposed quantum correlation framework. The computational complexity is $O(\log_2 N_m \log_2 N_k)$. The $m^2$ term can be written as:
\begin{equation}
m^2=\sum_p 2^{2p}m_p+2\sum_{p<q}2^{p+q}m_pm_q.
\end{equation}
The corresponding quantum circuit can be constructed as the single qubit phase gate $P(\eta 2^{2p})$ for each qubit $m_p$, and the controlled-phase gate $CP(2\eta 2^{p+q})$ for each pair of qubits $(m_p, m_q)$. The computational complexity is $O((\log_2 N_m)^2)$. If necessary, higher-order model polynomials can be implemented via multi-controlled phase gates. 

\item Quantum Fourier transform (QFT). Following fringe rotation, the $N_k$ sampling points within each FT period are transformed to the frequency domain via Fourier transform. In the quantum computing pipeline, this operation is implemented by applying a QFT to the first $n_k$ qubits of the station's quantum state. To avoid introducing additional notation, the expression $|m,k\rangle$ to represent this quantum state after the QFT is kept. At this stage, the state of $k$ transitions from a time-domain sampling point within each FT period to a frequency point within the same period.

\item Fractional Sampling Time Correction (FSTC). Due to the presence of the delay rate, FSTC value varies with FT periods. The fractional delay can be updated once per FT period based on the delay rate term $\dot{\tau}$:
\begin{equation}
   \tau (m)=\tau_0+\dot{\tau} m T_\FT.
\end{equation}
The corresponding phase modulation is:
\begin{align}
    \phi_\FSTC (m, k) & = 2\pi f_\bin\tau(m) k \nonumber \\
    & = 2\pi f_\bin\tau_0 k + 2\pi f_\bin\dot{\tau}T_\FT mk.
\end{align}
The phase modulation operator takes the form:
\begin{equation}\label{eq:P2}
P(\alpha, \beta): |m,k\rangle\rightarrow e^{j(\alpha k +\beta mk)} |m,k\rangle,
\end{equation}
with
\begin{align}
    \alpha & = 2\pi f_\bin\tau_0 \nonumber \\
    \beta  & = 2\pi f_\bin\dot{\tau}T_\FT. \nonumber
\end{align}
The implementation of above terms are identical to that of the corresponding terms in Step 2.

\item Integration. After completing the series of operations described above, the inner product of the quantum states of the two stations yields the integration of their cross-power spectrum along both the frequency and time dimensions:
\begin{equation}
S = \mathcal{M}(\langle \psi_B|\psi_A \rangle) = \sum_{m,k}a_{m,k}b^*_{m,k}
\end{equation}
\end{enumerate}

\subsection{Fringe Fitting}\label{sec:fringe_fitting}
In the current framework, it is only demonstrated that the fitting function for a series of trial delays implemented via quantum circuits. However, practical fringe fitting requires the consideration of both delay and delay rate. In this case, the phase modulation operator in Equation~(\ref{eq:Utau}) takes the form:
\begin{equation}
P(\alpha_d,~\beta_r):~|m, k\rangle \rightarrow e^{j(\alpha_d k + \beta_r m)}|m, k\rangle,
\end{equation}
with:
\begin{equation}
    \alpha_d=2\pi f_\mathrm{bin}\tau,~~\beta_r=2\pi f_0\dot{\tau}T_\FT.
\end{equation}
To improve the accuracy of the search, smaller step size is desirable, which increases the computational cost. It is worth noting that the Durr--Hoyer algorithm, which is the extension of the classical Grover search \citep{Grover1996}, can locate the maximum of a function defined on a discretized grid with complexity reduced from $O(N)$ to $O(\sqrt{N})$ \citep{DH1996}. However, the implementation of above approach requires several modifications to the current quantum algorithm. 

First of all, the set of trial delay and delay rate values must be loaded into quantum superposition state, which can be performed using the linear phase-modulation technique introduced in Section~\ref{sec:phase_mod}. Second, the cross-correlation computation must be encapsulated into an oracle, with the trial delay and delay rate values provided as a kind of ``argument'' to the quantum circuit, rather than constructing a separate circuit for each $\tau$ and $\dot{\tau}$ as done in the current design. Finally, since Grover-type algorithms require the oracle output to be coherent and reversible, the present Hadamard test based estimation cannot be used. A fully quantum-compatible method for amplitude evaluation or comparison must be developed instead. Most of these modifications are non-trivial and will require further investigation.

\section{Conclusions}\label{sec:conclusion}
This work demonstrates that VLBI data correlation is naturally compatible with quantum computation in several fundamental respects. On the one hand, a quantum state encoded in $n$ qubits spans a $2^n$-dimensional Hilbert space, enabling the representation of large volumes of raw baseband data using only $\log_2 N$ qubits. Such kind of quantum encoding provides a powerful form of data ``compression'' that is unachievable in classical systems. On the other hand, the basic operations in VLBI correlation and fringe fitting, including linear phase modulation, Fourier transform, and cross correlation, can be efficiently implemented via quantum circuits with lower computational complexity.

Based on these insights, a quantum VLBI data correlation pipeline using \texttt{Qiskit}, together with a corresponding classical pipeline for verification, is constructed. Using simulated baseband data, the full workflow, including fringe rotation, Fourier transform, FSTC, cross correlation, and fringe fitting, is tested. 
Within the margin of error, the classical and quantum results are consistent with each other, therefore validating the feasibility of the quantum approach. Moreover, it should be pointed out that current pipeline is primarily designed for geodetic VLBI. Further development is required to meet the requirements of data processing for radio astronomical observations.

Despite these promising results, several challenges remain before quantum computing can be applied to actual VLBI data correlation, particularly regarding quantum state preparation. Loading classical data into quantum superposition state via amplitude encoding is one of the major bottlenecks for quantum algorithms broadly, not merely for VLBI applications. Notably, VLBI raw baseband samples are typically quantized to 1 or 2 bits and exhibit repeated patterns, both of which can be investigated to reduce loading complexity. These characteristics suggest the feasibility of efficient state-preparation schemes. Furthermore, the current pipeline verification does not include cross spectrum integration and delay rate searching. However, it is demonstrated that these operations can be incorporated into the existing quantum computing framework, which will be validated in our future work.

As discussed in Section~\ref{sec:discussion}, both hardware implementation algorithm design and hardware implementations of quantum computing are not yet ready for practical deployment on operational radio interferometric arrays. However, Quantum computing offers substantial advantages over classical computation in efficient data representation and reduced computational complexity, which warrants further investigation, especially for large-scale radio interferometric arrays such as SKA. One of the main purpose of this work is to take geodetic VLBI as an example to inspire the community to explore quantum computing applications in radio interferometric signal processing and even broader astronomical research.

In general, this study demonstrates that quantum computing provides a promising framework for VLBI data correlation. With continued advances in quantum hardware, quantum-assisted algorithms may become a practical and powerful tool for next-generation interferometric arrays.

\vspace{6pt}



\vspace{4pt}
\noindent\textbf{Data Availability Statement:} The necessary data and code for the quantum pipeline are available and updated in the GitHub repository: \url{https://github.com/liulei/qcorr}.



\vspace{6pt}
\noindent\textbf{Abbreviations}\\
The following abbreviations are used in this manuscript:\\
\noindent 
\begin{tabular}{@{}ll}
FFT & Fast Fourier Transform\\
FSTC & Fractional Sample Time Correction\\
FT & Fourier Transform \\
HOPS & Haystack Observatory Postprocessing System\\
IF & Intermediate Frequency \\
NISQ & Noisy Intermediate-Scale Quantum \\
QC & Quantum Computing \\
QFT & Quantum Fourier Transform\\
SKA & Square kilometre Array\\
SNR & Signal-to-Noise Ratio\\
VLBI & Very Long Baseline Interferometry
\end{tabular}

\appendix
\section{Hadamard Test}\label{sec:Hadamard}
In quantum computing, the Hadamard Test is a classical algorithm used to measure the expectation value $\langle\psi|U|\psi\rangle$ (including both the real and imaginary parts) of a quantum state $|\psi\rangle$ under the operator $U$.

\begin{itemize}
\item Real part ($H \rightarrow \text{controlled-}U \rightarrow H$): \\
First of all, an ancilla initialized to $|0\rangle$. The initial state is given by:
\begin{equation}
    |0\rangle |\psi\rangle.
\end{equation}
Applying a Hadamard gate to the ancilla:
\begin{equation}\label{eq:Re_H1}
\frac{1}{\sqrt{2}}(|0\rangle + |1\rangle) |\psi\rangle.
\end{equation}
Applying a controlled-$U$ gate:
\begin{equation}\label{eq:Re_cU}
\frac{1}{\sqrt{2}}|0\rangle|\psi\rangle + \frac{1}{\sqrt{2}}|1\rangle U|\psi\rangle.
\end{equation}
Applying the second Hadamard gate to the ancilla. Noting that $H|0\rangle = \frac{1}{\sqrt{2}}(|0\rangle+|1\rangle)$ and $H|1\rangle = \frac{1}{\sqrt{2}}(|0\rangle-|1\rangle)$, one obtains:
\begin{equation}
 \frac{1}{2}|0\rangle(|\psi\rangle + U|\psi\rangle) + \frac{1}{2}|1\rangle(|\psi\rangle - U|\psi\rangle).
\end{equation}
The probability of measuring the ancilla in state $|0\rangle$ is given by:
\begin{align}
P_0 = \left| \frac{|\psi\rangle + U|\psi\rangle}{2} \right|^2 &= \frac{1}{4} \left( \langle\psi| + \langle\psi|U^\dagger \right) \left( |\psi\rangle + U|\psi\rangle \right)\nonumber \\ 
&= \frac{1}{4} \left( \langle\psi|\psi\rangle + \langle\psi|U|\psi\rangle \right.\nonumber \\
& ~~~~ + \left. \langle\psi|U^\dagger|\psi\rangle + \langle\psi|U^\dagger U|\psi\rangle \right).
\end{align}
Noting that $\langle\psi|\psi\rangle = 1$, $U$ is a unitary operator ($U^\dagger U = I$), $\langle\psi|U^\dagger|\psi\rangle$ is the complex conjugate of $\langle\psi|U|\psi\rangle$, one obtains:
\begin{equation}
P_0 = \frac{1}{2}(1 + \mathrm{Re}\langle\psi|U|\psi\rangle),
\end{equation}
then:
\begin{equation}
\mathrm{Re}\langle\psi|U|\psi\rangle=2P_0-1.
\end{equation}

\item Imaginary part ($H \rightarrow S^\dagger \rightarrow \text{controlled-}U \rightarrow H$): \\
The main difference between measuring the imaginary and real parts is the insertion of an $S^\dagger$ gate after the first Hadamard gate (\ref{eq:Re_H1}) and prior to the controlled-$U$ gate (\ref{eq:Re_cU}). Since $S^\dagger|0\rangle = |0\rangle$ and $S^\dagger|1\rangle = -i|1\rangle$, one obtains:
\begin{equation}
\frac{1}{\sqrt{2}}(|0\rangle - i|1\rangle) |\psi\rangle.
\end{equation}
The subsequent operations and measurement are identical to those of the real part. Finally one obtains:
\begin{equation}
\mathrm{Im}\langle\psi|U|\psi\rangle=2P_0-1.
\end{equation}

\end{itemize}

In the actual implementation, after constructing the quantum circuit with an ancilla and the controlled-$U$ operator, the circuit is executed on either a quantum simulator or quantum hardware. In each run (shot), the ancilla is measured in the computational basis, yielding a classical outcome of either 0 or 1. By repeating the circuit $N_\mathrm{shots}$ times, one obtains the measurement frequencies $N_0$. $P_0=N_0 / N_\mathrm{shots}$ is derived in this way.

\section{Estimation of Maximum Allowable Integration Time}\label{sec:Tmax}
The VLBI delay model is written as:
\begin{equation}
    \tau(t) = a_0+ a_1 t + a_2t^2 + \dots
\end{equation}
In fringe rotation, the phase to be rotated at time $t$ is:
\begin{equation}
    \phi(t)=2\pi f_0 \tau(t).
\end{equation}
If the evaluation time for the fringe rotation coefficients is chosen at the midpoint of the total integration time $T$, the phase error caused by the $n$-th term is given by:
\begin{equation}
    \Delta\phi_n = 2\pi f_0 a_n (T/2)^n.
\end{equation}
For the given phase error $\epsilon$, the maximum allowable integration time assuming that $a_n$ term is neglected is:
\begin{equation}
    T_{n,\mathrm{FR}}=2 \left(\frac{\epsilon}{2\pi f_0 a_n}\right)^{1/n}.
\end{equation}

The same estimation method can be used for FSTC. The maximum phase error across the entire bandwidth $B$ is given by:
\begin{equation}
    \Delta\phi_n = 2\pi B a_n (T/2)^n.
\end{equation}
The corresponding maximum allowable integration time assuming that $a_n$ term is neglected is:
\begin{equation}
    T_{n,\mathrm{FSTC}}=2\left(\frac{\epsilon}{2\pi B a_n}\right)^{1/n}.
\end{equation}


\begin{thebibliography}{999}
\bibitem[Feynman(1982)]{Feynman1982} Feynman, R.P. Simulating physics with computers. \textit{Int. J. Theor. Phys.} \textbf{1982}, \textit{21}, 467. \url{https://doi.org/10.1007/BF02650179}

\bibitem[Deutsch(1985)]{Deutsch1985} Deutsch, D. Quantum Theory, the Church-Turing Principle and the Universal Quantum Computer. \textit{Proc. R. Soc. Lond. A} \textbf{1985}, \textit{400}, 97. \url{https://doi.org/10.1098/rspa.1985.0070}

\bibitem[Shor(1994)]{Shor1994} Shor, P.W. Algorithms for quantum computation: discrete logarithms and factoring. In Proceedings of the 35th IEEE Symposium on Foundations of Computer Science, Santa Fe, NM, USA, 1994; p. 124. \url{https://doi.org/10.1109/SFCS.1994.365700}

\bibitem[Nielsen \& Chuang(2010)]{QCQI} Nielsen, M.A.; Chuang, I.L. \textit{Quantum Computation and Quantum Information}, 2nd ed.; Cambridge University Press: Cambridge, UK, 2010.

\bibitem[Arute et al.(2019)]{google} Arute, F.; Arya, K.; Babbush, R.; et al. Quantum supremacy using a programmable superconducting processor. \textit{Nature} \textbf{2019}, \textit{574}, 505. \url{https://doi.org/10.1038/s41586-019-1666-5}

\bibitem[Zhong et al.(2020)]{jiuzhang} Zhong, H.-S.; Wang, H.; Deng, Y.-H.; et al. Quantum computational advantage using photons. \textit{Science} \textbf{2020}, \textit{370}, 1460. \url{https://doi.org/10.1126/science.abe8770}

\bibitem[Au-Yeung et al.(2024)]{AY2024} Au-Yeung, R.; Camino, B.; Rathore, O.; et al. Quantum algorithms for scientific computing. \textit{Rep. Prog. Phys.} \textbf{2024}, \textit{87}, 116001. \url{https://doi.org/10.1088/1361-6633/ad85f0}

\bibitem[Thompson et al.(2017)]{VLBI} Thompson, A.R.; Moran, J.M.; Swenson, G.W. \textit{Interferometry and Synthesis in Radio Astronomy}, 3rd ed.; Springer: Cham, Switzerland, 2017. \url{https://doi.org/10.1007/978-3-319-44431-4}

\bibitem[Event Horizon Telescope Collaboration(2019)]{EHT} Event Horizon Telescope Collaboration; Akiyama, K.; Alberdi, A.; et al. First M87 Event Horizon Telescope Results. I. The Shadow of the Supermassive Black Hole. \textit{ApJL} \textbf{2019}, \textit{875}, L1.

\bibitem[Haas et al.(2015)]{VGOS} Haas, R.; Nothnagel, A.; Petrachenko, B. VGOS: The VLBI Global Observing System of the IVS. \textit{IAU Gen. Assem.} \textbf{2015}, \textit{29}, 2257511.

\bibitem[Duev et al.(2012)]{Duev2012} Duev, D.A.; Molera Calv{\'e}s, G.; Pogrebenko, S.V.; et al. Spacecraft VLBI and Doppler tracking: algorithms and implementation. \textit{A\&A} \textbf{2012}, \textit{541}, A43.

\bibitem[Whitney(2007)]{Whitney2007} Whitney, A.R. The Mark 5 VLBI Data System. In Proceedings of the 18th European VLBI for Geodesy and Astrometry Work Meeting, 2007; Volume 79, p. 33.

\bibitem[Zhu et al.(2016)]{Zhu2016} Zhu, R.; Wu, Y.; Li, J. Progress on the VLBI Data Acquisition System of SHAO. In \textit{New Horizons with VGOS}; 2016; p. 163.

\bibitem[Zheng et al.(2010)]{CVNCORR} Zheng, W.; Quan, Y.; Shu, F.; et al. The Software Correlator of the Chinese VLBI Network. In Proceedings of the Sixth International VLBI Service for Geodesy and Astronomy 2010 General Meeting, 2010; p. 157.

\bibitem[Deller et al.(2011)]{difx} Deller, A.T.; Brisken, W.F.; Phillips, C.J.; et al. DiFX-2: A more flexible, efficient, robust, and powerful software correlator. \textit{PASP} \textbf{2011}, \textit{123}, 275. \url{https://doi.org/10.1086/658907}

\bibitem[Dewdney et al.(2009)]{SKA} Dewdney, P.E.; Hall, P.J.; Schilizzi, R.T.; et al. The square kilometre array. \textit{Proc. IEEE} \textbf{2009}, \textit{97}, 1482. \url{https://doi.org/10.1109/JPROC.2009.2021005}

\bibitem[Kondo \& Takefuji(2016)]{Kondo2016} Kondo, T.; Takefuji, K. An algorithm of wideband bandwidth synthesis for geodetic VLBI. \textit{Radio Sci.} \textbf{2016}, \textit{51}, 1686. \url{https://doi.org/10.1002/2016RS006070}

\bibitem[Javadi-Abhari et al.(2024)]{Qiskit} Javadi-Abhari, A.; Treinish, M.; Krsulich, K.; et al. Quantum computing with Qiskit. \textit{arXiv} \textbf{2024}, arXiv:2405.08810. \url{https://doi.org/10.48550/arXiv.2405.08810}

\bibitem[Whitney et al.(2022)]{HOPS} Whitney, A.R.; Cappallo, R.; Aldrich, W.; et al. Haystack Observatory Postprocessing System. \textit{Astrophys. Source Code Libr.} \textbf{2022}, ascl:2205.019.

\bibitem[Grover(1996)]{Grover1996} Grover, L.K. A fast quantum mechanical algorithm for database search. \textit{arXiv} \textbf{1996}, arXiv:quant-ph/9605043. \url{https://doi.org/10.48550/arXiv.quant-ph/9605043}

\bibitem[Durr \& Hoyer(1996)]{DH1996} Durr, C.; Hoyer, P. A Quantum Algorithm for Finding the Minimum. \textit{arXiv} \textbf{1996}, arXiv:quant-ph/9607014. \url{https://doi.org/10.48550/arXiv.quant-ph/9607014}
\end{thebibliography}
\end{document}